\documentclass[aps,prb,onecolumn,nofootinbib,citeautoscript,10pt]{revtex4-2}

\usepackage{amsmath,amssymb} 
\usepackage{comment}

\usepackage[tight]{subfigure} 

\usepackage[dvipsnames]{xcolor} 
\usepackage[papersize={8.5in,11in}]{geometry}
\usepackage[colorlinks=true]{hyperref}
\hypersetup{
    bookmarks=true,         
    unicode=false,          
    pdftoolbar=true,        
    pdfmenubar=true,        
    pdffitwindow=false,     
    pdfstartview={FitH},    
    pdfkeywords={keyword1} {key2} {key3}, 
    pdfnewwindow=true,      
    colorlinks=true,       
    linkcolor=magenta, 
    citecolor=blue,        
    filecolor=magenta,      
    urlcolor=blue           
} 

\usepackage{dcolumn}
\usepackage{color}
\usepackage{amssymb,amsmath}
\usepackage{tabularx,graphicx}
\usepackage{epstopdf}
\usepackage{latexsym}
\usepackage{colortbl}
\usepackage{psfrag}
\usepackage{bbm,bm,array,physics}
\usepackage{dsfont}
\usepackage{float, mathrsfs}

\def \nn{\nonumber \\}
\def\*#1{\boldsymbol{#1}} 

\begin{document}

\title{Disentangling contributions to longitudinal magnetoconductivity for Kramers-Weyl nodes}

\author{Ipsita Mandal}
\email{ipsita.mandal@snu.edu.in}

\affiliation{Department of Physics, Shiv Nadar Institution of Eminence (SNIoE), Gautam Buddha Nagar, Uttar Pradesh 201314, India}

\begin{abstract} 
We set out to compute the longitudinal magnetoconductivity for an isolated and isotropic Kramers-Weyl node (KWN), existing in chiral crystals, which forms an exotic cousin of the conventional Weyl nodes resulting from band-inversions. The peculiarities of KWNs are many, the principal one being the presence of two concentric Fermi surfaces at any positive chemical potential ($\mu$) with respect to the nodal point. This is caused by a dominant quadratic-in-momentum dispersion, with the linear-in-momentum Dirac- or Weyl-like terms relegated to a secondary status. In a KWN, the chirally-conjugate node typically serves as a mere doppelgänger, being significantly separated in energy. Hence, when $\mu$ is set near such a node, the signatures of a lone node are probed in the transport-measurements. The intrinsic topological quantities in the forms of Berry curvature and orbital magnetic moment contribute to the linear response, which we determine by exactly solving the semiclassical Bolzmann equations. Another crucial feature is that the two bands at the same KWN node carry actual spin-quantum numbers, thus providing an additional coupling to an external magnetic field ($\boldsymbol B$), and affecting the conductivity. We take this into account as well, and demonstrate that it causes a linear-in-$B$ dependence, on top of the usual $B^2$-dependence.
\end{abstract}

\maketitle

\tableofcontents


\section{Introduction}

The advent of the discovery of materials like the Weyl semimetals (WSMs) ~\cite{burkov11_Weyl, armitage_review, yan17_topological} has opened up the rewarding exercise of identifying the signatures of topological properties in transport properties, emerging from the Brillouin zone (BZ) of three-dimensional (3d) semimetals. From group-theoretical considerations, it has been known for quite some time that there exists a class of materials, called the chiral crystals, which host the so-called \textit{Kramers-Weyl nodes} (KWNs), referring to a Weyl node pinned at a point with time-reversal-invariant momentum (TRIM) \cite{bernevig, chang2018, hasan-review} (although free to move in energy). In a stark contrast with the conventional band-inversion Weyl nodes, these are not allowed to deviate from the TRIMs in the momentum space, under the action of small perturbations \cite{bernevig, chang2018}. The chiral crystals are named so because they possess a sense of handedness (i.e., chirality) due to the lack of any roto-inversion symmetries. In particular, for symmorphic chiral crystals, every Kramers-pair of bands at every TRIM is guaranteed to host a Weyl node. On the other hand, in non-symmorphic chiral crystals, this holds only for a subset of TRIMs, which does always include the $\Gamma$-point (i.e., the centre of the BZ). The KWNs arise in non-magnetic chiral crystals due to the presence of spin–orbit coupling (SOC), without a band-inversion involved. Basically, SOC acts to annihilate the band-inversion Weyl nodes, while creating KWNs, allowing the coupling between the spin and momentum degrees of freedom.

The hallmark feature of a twofold node (e.g., in WSMs) or a multifold node in a 3d semimetal is its nontrivial topology \cite{xiao_review, sundaram99_wavepacket, graf-Nband} in the momentum space, resulting from the Berry phases \cite{berry}, which shows up in the form of topological properties like the Berry curvature (BC) and the orbital magnetic moment (OMM). These, in turn, cause exotic characteristics in the response, that are detectable in transport-experiments \cite{timm, ips_rahul_ph_strain, rahul-jpcm, ips-kush-review, claudia-multifold, ips-ruiz, ips-rsw-ph, ips-tilted, ips-shreya, ips-spin1-ph}. The analogy of the BC to a magnetic field leads to the picture of nonzero BC monopoles sitting at the nodal points \cite{fuchs-review, polash-review}, because they are the sources and sinks of the BC-flux field. Mathematically, for the case of twofold nodal-degeneracies, the monopoles' charges also represent the Chern numbers of the closed surfaces enclosing the nodal singularities. Conventionally, we designate it as the Chern number ($\mathcal C$) of the lower-energy band, while the higher-energy one is assigned a value equalling $ - \, \mathcal C$. The sign of $\mathcal C$ is pictured as the chirality ($\chi$) of the node, associating a sense of \textit{handedness or chirality} with the quasiparticles occupying the bands meeting there. Accordingly, they are called \textit{right-handed} or \textit{left-handed}, depending on whether $\chi = 1$ or $\chi = -1$.

In crystal lattices, it is essential to satisfy the Nielsen-Ninomiya theorem \cite{nielsen81_no}, which is physically reflected by the fact that the Weyl nodes (with opposite chiralities) must always exist in pairs in the entire BZ. For the spinless WSMs, arising in achiral crystals (e.g., TaAs family \cite{lv_Weyl}), such conjugate partners are typically (almost) degenerate in energy, due to the presence of mirror or other roto-inversion symmetries. Therein, charge-pumping is an important internode phenomenon, in the presence of external electric ($\boldsymbol E $) and magnetic ($\boldsymbol B $) fields, having nonvanishing collinear components. This is an embodiment of the chiral anomaly in the arena of condensed matter physics \cite{chiral_ABJ, son13_chiral, li_nmr17, ips-internode}, where the analog of the well-known concept of spacetime chirality (of relativistic fermions) is realised in the momentum space. On the contrary, the oppositely-charged chiral nodes in chiral crystals need not be degenerate in energy, because the conjugate nodes are not related by crystal symmetries. In fact, they are observed to have discernible separations in energy and momenta, with an isolated KWN located at an intrinsic chemical potential \cite{law-ksm, chang2018}. As such, internode-scattering-induced charge-pumping becomes unimportant, while enhancing other phenomena like quantised circular photogalvanic effect \cite{bernevig, chang2018, ni2021_giant, moore18_optical, guo23_light,kozii, ips_cpge} and circular dichroism \cite{ips-cd1, ips_cd}. Another crucial distinction is that, in a band-inversion WSM, the pseudospin of the two-component fermions represents a quantum number \textit{analogous to} the (but not the actual) spacetime spin. On the other hand, a two-component spinor, representing the Kramers–Weyl quasiparticles, is written in the space of the electron-spin matrices ($\boldsymbol \sigma$), in the limit when the energy-scale of the SOC is much smaller than the interband-separation of the zero-SOC scenario \cite{chang2018}. Consequently, measurements of spin-polarizations can directly capture the Chern numbers of the associated bands. Last but not the least, a linear-in-momentum term, accompanying the identity matrix (in the operator space for the spinors), is not allowed in their effective Hamiltonians, because of the presence of the time-reversal symmetry. Hence, tilting is completely ruled out for the KWNs, although it is a generic feature of the conventional Weyl cones \cite{yadav23_magneto, staalhammar20_magneto, rahul-jpcm, ips-ruiz, ips-tilted}, due to the less restrictions imposed by the low-symmetry points of the BZ.

Experimentally, the smoking-gun properties of KWNs have been observed in materials like $\beta$-Ag$_2$Se \cite{long-ksm}, which crystallises in a nonsymmorphic chiral form. In this paper, we will consider isotropic KWNs, which arises at a node harbouring the high-symmetry cubic point group, $\lbrace \rm T, O \rbrace $. The candidate materials include K$_2$Sn$_2$O$_3$ \cite{chang2018}, RhSi \cite{sanchez}, CoSi \cite{sanchez}, AlPt \cite{schroter}, and PtGa \cite{mengyu}.

\begin{figure*}[t]
\subfigure[]{\includegraphics[width = 0.25 \textwidth]{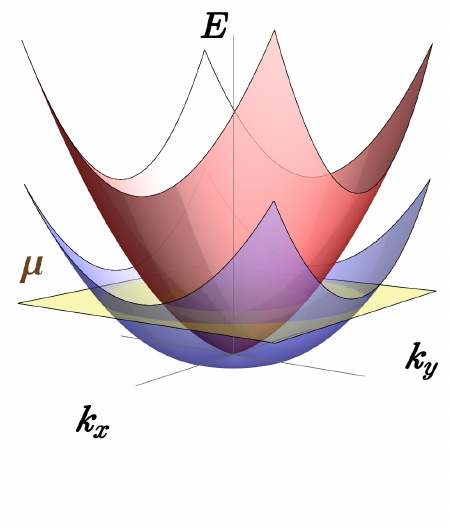}} \hspace{3 cm}
\subfigure[]{\includegraphics[width = 0.32 \textwidth]{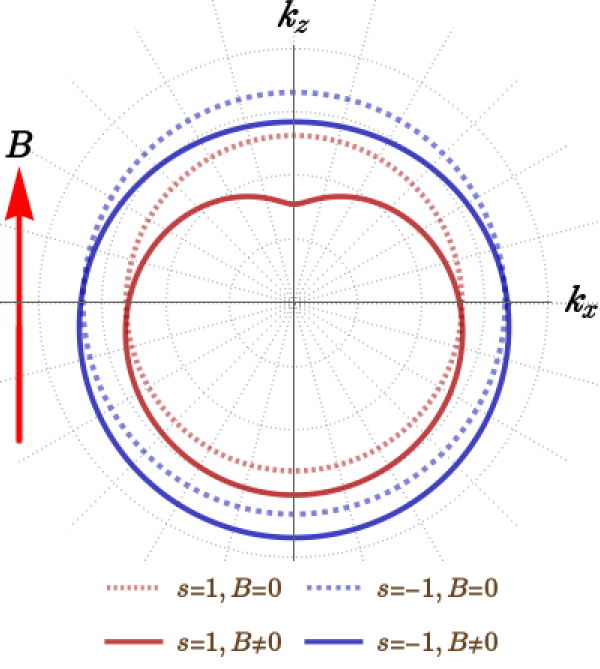}}
\caption{\label{figdis}Energy bands for $s=1$ (light red) and $s=-1$ (light blue): (a) Bare dispersion ($E $) of the two bands at an isotropic Kramers-Weyl node against the $k_x k_y$-plane (or, equivalently, $k_y k_z$- and $k_z k_x$-planes). The yellow plane depicts a positive chemical potential ($\mu$) cutting the bands, giving rise to two concentric Fermi surfaces. (b) Schematics of the Fermi-surface projections in the the $k_z k_x$-plane, when a nonzero $\boldsymbol B$ (red arrow) is applied along the $z$-axis. To provide an eye-estimate, the dotted curves show the unperturbed projections in the absence of an external magnetic field.}
\end{figure*}

In our recent work, we addressed the question of computing longitudinal magnetoconductivity in systems where multiple bands contribute to transport at a particular node \cite{ips-exact-spin1}. In this paper, we address a similar question where an isotropic KWN (say, at the $\Gamma$-point) is the system under consideration. As pointed out earlier, a single node will be relevant here, when we tune the external chemical potential to lie near the intrinsic energy of the concerned nodal point, as the chirally-conjugate node is well-separated in energy and momentum (see, for example, Refs. \cite{titus-ksm, law-ksm}). In other words, the energy-separation between the $\chi =\pm $ members of the pair is much larger than the temperature-scale of the experiments. Now, the low-energy effective Hamiltonian [see Eq.~\eqref{eqham} discussed later] shows that the two concentric Fermi surfaces coexist at a positive chemical potential ($\mu$) compared to the nodal point \cite{long-ksm}. The schematics of the dispersion is depicted in Fig.~\ref{figdis}. This is brought about by the dominance of the quadratic-in-momentum term in the dispersion over the linear terms. Basically, instead of one positive and one negative bands of a conventional Weyl cone, here we end up with two positive-energy bands, both of which will participate in transport for $\mu > 0 $. Thus, the behaviour is somewhat similar to the multifold (or higher-pseudospin) nodes, and the nontrivial topological properties of each band is expected to influence the linear response with the elements of BC and OMM \cite{ips-ruiz, ips-rsw-ph, ips-spin1-ph, ips-nl-ph, ips-exact-spin1}. In the semiclassical picture, a Bloch electron is modelled by a wave-packet in a Bloch band, which is found to rotate about its center of mass in general, yielding an intrinsic magnetic moment dubbed as the OMM, which is topological in origin \cite{chang-niu, sundaram99_wavepacket, xiao06_berry, xiao_review,xiao07_valley}. In the presence of a magnetic field, the electronic band-structure energy acquires a correction term from this intrinsic orbital moment. This invariably generates a modification of the Fermi surface, making it anisotropic, as shown in Fig.~\ref{figdis} (b). Hence, it brings about the possibility of the OMM generating nontrivial terms compared to the case when it is neglected. Now, whether it gives rise to a nonzero term or not depends on whether it can generate a nonzero intergral, similar to the BC part. This can be roughly determined by looking at whether the resulting integrand contains odd or even powers of the components of the momentum [see, for example, the discussions around Eq.~\eqref{def_sig_long2}].

It is interesting to note that, since the Chern numbers of the two Fermi surfaces are exactly opposite, there are no Fermi arcs visible on any surface-BZ. While the BC profiles of the two bands differ by an overall sign, the OMM-vector fields are identical. 
An additional ingredient, which one does not have to consider for the pseudospin-space spinors, is the energy modification caused by the spin magnetic moment (SMM). Needless to say, it comes with opposite signs for the pair of bands \cite{law-ksm}. By coupling to the applied magnetic field, the BC, OMM, and SMM affect the conductivity, which we aim to compute here. Equipped with the versatile tool of Boltzmann's semiclassical transport equations, we will venture into computing the longitudinal conductivity exactly, arising from applying collinear electromagnetic fields, $\boldsymbol   E  \parallel \boldsymbol B $. The results will be limited to the realm of weak/nonquantising magnetic fields (in the sense that the Landau-level splitting is too tiny), but by going beyond the relaxation-time approximation. We will follow the procedure developed in Ref. \cite{timm} for achiral WSMs.

The paper is organised as follows: In Sec.~\ref{secmodel}, we describe the explicit form of the low-energy effective Hamiltonian in the vicinity of a KWN node. Sec.~\ref{secboltz} is devoted to deriving the equations leading to the final values of the longitudinal magnetoconductivity. The results are discussed in Sec.~\ref{secres} therein, illustrated by representative plots. For the sake of completeness, in Sec.~\ref{secsig}, we derive the conductivity in a relaxation-time approximation. Finally, we conclude with a summary and outlook in Sec.~\ref{secsum}. In all our expressions, we will resort to the simplification offered by using the natural units, which involves setting the reduced Planck's constant ($\hbar $), the speed of light ($c$), and the Boltzmann constant ($k_B $) to unity. Additionally, the electric charge has no units. Despite the fact that the magnitude ($e$) of a single unit of electrons' charges is unity, we will retain it in our expressions just as a matter of book-keeping.

\section{Model}
\label{secmodel}

Our focus is on isotropic and isolated KWNs, which appear at the $\Gamma$-points of chiral crystals, stabilised by the point groups isomorphic to $\lbrace \rm T, O \rbrace $ \cite{law-ksm}. The low-energy effective Hamiltonian in the vicinity of such a node, obtained through the standard $\mathbf k \cdot \mathbf p$ method, is captured by
\begin{align} 
\label{eqham}
\mathcal{H}  ( \boldsymbol k) & = 
\frac{k^2} {2\, m}  \; \mathbb{I}_{2\times 2}
+ v_0 \, \boldsymbol k \cdot \boldsymbol{\sigma}  \,,
\quad k =\sqrt{k_x^2 +k_y^2 +k_z^2} \,.
\end{align}
Here, $\boldsymbol  \sigma  = \lbrace \sigma_x, \,\sigma_y, \, \sigma_z \rbrace$ represents the vectorial spin-1/2 operator, comprising the three components of the Pauli matrices, $m$ is the effective mass, and $v_0 $ is the isotropic Fermi-velocity for the linear-in-momentum part.
The energy eigenvalues of the Hamiltonian are given by
 \begin{align} 
\label{eigenvalues}
\varepsilon^s  ({ \boldsymbol k}) =  c\,  k^2 + s \,v_0 \,  k  \,, \quad
s =\pm 1 \,, \quad c =\frac{1}{2 \, m} \,,
\end{align}
which can be visualised in Fig.~\ref{figdis}(a).
We note that, for the conventional case of band-inversion WSMs, $c$ goes to zero, such that $s=-1$ is a negative-energy band. For the KWNs, the $ s \,v_0\,  k $ terms being subdominant to the $ c\,  k^2$ part, both the bands contribute to transport for $\mu > 0$.
The group-velocity of the quasiparticles, occupying the band with index $s$, is given by
\begin{align}
{\boldsymbol  v}^s ( \boldsymbol k) 
\equiv \nabla_{\boldsymbol k} \varepsilon^s   (\boldsymbol k)
=  \frac{ 2 \, c\, k + s\, v_0  } {  k }  
\left \lbrace k_x,\,k_y, \, k_z \right \rbrace .
\end{align}

Exploiting the isotropy of $\mathcal H $, we will switch to the convenience of using the spherical-polar coordinates, such that
\begin{align}
\label{eqcyln} 
k_x =  k \sin \theta  \cos \phi \,, \quad
k_y =   k \sin \theta  \sin \phi \,, \quad k_z = k \cos \theta\,,
\end{align}
where $ k \in [0, \infty )$, $\phi \in [0, 2 \pi )$, and $\theta \in [0, \pi ]$. 
A set of orthonormal eigenvectors, $ \lbrace \psi_{s} (\boldsymbol k) \rbrace $, can be expressed as follows:
\begin{align}
\label{eqev}
 \begin{bmatrix}
 \cos \big (\frac{\theta }{2} \big) &
e^{i \,\phi } \sin \big (\frac{\theta }{2} \big)
\end{bmatrix}^{\rm T}
 \text{ for }  s = 1 \,,\qquad
 \begin{bmatrix}
- \, \sin \big (\frac{\theta }{2} \big)
&  e^{i \, \phi } \cos \big (\frac{\theta }{2} \big)
\end{bmatrix}^{\rm T}
\text{ for }  s = -1 \,.
\end{align} 
Here, we note that, since the $c \, k^2$ term accompanies the identity matrix, the spinor-structure of the eigenvectors is the same as for a conventional Weyl node. Therefore, the topological properties like the BC and the OMM are identical to the WSM case, whose expressions are given below.

The SMM for the two bands can be written as
\begin{align}
{\boldsymbol S}^s = \frac{ -\, e \,g\,\mu_B \,\boldsymbol \sigma} {2} \, ,
\end{align}
where, $\mu_B $ is the Bohr magneton and $g$ is the Land\'e g-factor, which we shall set to two in our calculations.
Overall, we approximate the value of $ { e \,g\,\mu_B } /{2} $ by $3 \times 10^{-7}$ eV$^{-1}$ in our numerics.
For the $s^{\rm th}$ band, the vector fields for the BC and the OMM are obtained using \cite{xiao_review,xiao07_valley}
\begin{align} 
& {\boldsymbol \Omega}^s ( \boldsymbol k)  = 
    i  \left[ \nabla_{ \boldsymbol k}  \psi_{s}({ \boldsymbol k}) \right ]^\dagger 
    \cross  \left [ \nabla_{ \boldsymbol k}  \psi_{s}({ \boldsymbol k}) \right ]
\text{and } 
{\boldsymbol {m}}^s ( \boldsymbol k) = 
\frac{  -\,i \, e} {2 } \,
 \left[ \nabla_{ \boldsymbol k} \psi_{s} ({ \boldsymbol k}) \right ]^\dagger 
 \cross
\Big [
\left \lbrace \mathcal{H} ({ \boldsymbol k}) -\varepsilon^s
({ \boldsymbol k}) 
\right \rbrace
\left \lbrace \boldsymbol \nabla_{ \boldsymbol k} \psi_{s}({ \boldsymbol k})
 \right \rbrace \Big  ] ,
\end{align}
respectively. For our two-band system, the expressions above simplify to \cite{fuchs-review}
\begin{align} 
\label{eqomm}
\Omega^s_i ( \boldsymbol  k) = \frac{  (-1)^{s} \,  
\epsilon_{ijl}}
 {4\,| \boldsymbol  k |^3} \, \boldsymbol k  \cdot
 \left[   \partial_{k_j} \boldsymbol k \cross  
 \partial_{k_l } \boldsymbol k \right ] \text{ and }
\boldsymbol m^s ( \boldsymbol  k) = 
 e\left( s \, v_0 \, k\right) \boldsymbol \Omega \,.
\end{align}
On evaluating the expressions in Eq.~\eqref{eqomm} for $\mathcal{H} ( \boldsymbol  k) $, we get
\begin{align}
\label{eqbcomm}
{\boldsymbol \Omega}^s( \boldsymbol k) = 
\frac{ -\, s  } { 2\, k^3 }
 \left \lbrace k_x, \, k_y, \,  k_z \right \rbrace, \quad 
{\boldsymbol m}^s( \boldsymbol k) = 
\frac{ -\,  e\, v_0 } {2 \, k^2 } 
 \left \lbrace k_x, \, k_y, \,  k_z \right \rbrace .
\end{align}
Clearly, the BC changes sign with $s$, but the OMM does not. Hence, in what follows, we will remove the unnecessary superscript ``$s$'' from ${\boldsymbol{m}}^s ({ \boldsymbol  k})$.

\section{Conductivity}
\label{secboltz}

In this section, we will review the methodology to compute the electric conductivity, applying the machinery of semiclassical Boltzmann equations \cite{mermin, sundaram99_wavepacket, li2023_planar, ips-kush-review, ips-rsw-ph, ips-shreya, timm, ips-exact-spin1}. 
We primarily follow and build up on the procedure developed in Ref.~\cite{timm} for obtaining the exact solutions.

\subsection{Influence of spin and topology}

We start by outlining how the OMM and the SMM modify the effective dispersion of each band, when we apply an external magnetic field. Both of them couple to $\boldsymbol B $, generating the corrections of \cite{xiao_review}
\begin{align}
\label{eqmodi}
& \varepsilon^s_{(\sigma)}  
= -\, {\boldsymbol B} \cdot {\boldsymbol S}_s \text{ and }
\varepsilon_{(m)}   (\boldsymbol k) 
= -\, {\boldsymbol B} \cdot \boldsymbol{m} (\boldsymbol k) \,,
\end{align}
respectively. Clearly, the $s$-dependence of $\varepsilon^s_{(\sigma)} $ changes the effective chemical potential at the bands as $\mu \rightarrow \mu - \varepsilon^s_{(\sigma)} $, acting in an opposite fashion for the two bands. On the other hand, the OMM distorts the Fermi surfaces along the direction of $\boldsymbol B$, as schematically depicted in Fig.~\ref{figdis}(b). In addition to affecting the Fermi-Dirac-distribution functions, it causes modifications of the group-velocities via adding the following term to ${\boldsymbol  v}^s ( \boldsymbol k)$:
\begin{align}
 {\boldsymbol   v}_{(m)} ({\boldsymbol k} ) \equiv 
 \nabla_{\boldsymbol k} \varepsilon_{(m)}   (\boldsymbol k) \,.
\end{align}
We would like to emphasize that, in twofold and multifold nodal points carrying pseudospin representations, an SMM contribution does not arise and, hence, did not appear in our earlier studies of magnetotransport \cite{rahul-jpcm, ips-ruiz, ips-tilted, ips-internode, ips-rsw-ph, ips-spin1-ph, ips-exact-spin1}. Furthermore, the effects of both the SMM and OMM have been neglected in the literature covering KWNs \cite{titus-ksm, law-ksm}. No wonder that the SMM adds an extra layer to the exotic signatures expected from linear response.

A nonzero BC manifests itself primarily via modifying the phase-space volume element, incorporated through the factor of \cite{mermin, sundaram99_wavepacket, li2023_planar, ips-kush-review, ips-rsw-ph, ips-shreya}
\begin{align}
{\mathcal D}_{s}  (\boldsymbol k) = \left [1 
+ e \,  \left \lbrace 
{\boldsymbol B} \cdot \boldsymbol{\Omega}^s  (\boldsymbol k)
\right \rbrace  \right ]^{-1}.
\end{align}
This, in turn, it affects the explicit forms of the Hamilton's equations, which we discuss below.

\subsection{Kinetic equations driven by electromagnetic fields}

The equilibrium Fermi-Dirac distribution,
\begin{align}
\label{eqdist}
f_0 \big (\xi_s (\boldsymbol k) , \mu, T \big )
= \left[ 1 + \exp \left \lbrace  
\left(  \xi_s (\boldsymbol k)-\mu \right) /T  \right \rbrace \right ]^{-1}, 
\end{align}
where $T $ is the temperature, contains the overall energy,
\begin{align}
\xi_s ({ \boldsymbol k}) =
 {\tilde \varepsilon}_s  ({ \boldsymbol k})  
+ \varepsilon_{(m)}   (\boldsymbol k) \,, \quad
 {\tilde \varepsilon}_s  ({ \boldsymbol k})=
 \varepsilon^s  ({ \boldsymbol k}) +  \frac{ e \,g\,\mu_B \,s \, B} {2}\,.
\end{align}
While using $f_0$ in various equations, we will be suppressing its $\mu$- and $ T $-dependence for uncluttering of notations. Furthermore, in what follows, we will perform our calculations by setting $ T = 0$ and considering $\mu > 0 $.

The Hamilton's equations of motion for the electronic quasiparticles, occupying a given Bloch band, are described by \cite{mermin, sundaram99_wavepacket, li2023_planar, ips-kush-review, ips-rsw-ph, ips-shreya}
\begin{align}
\label{eqrkdot}
\dot {\boldsymbol r} & = \nabla_{\boldsymbol k} \, \xi_s  
- \dot{\boldsymbol k} \, \cross \, \boldsymbol \Omega^s  (\boldsymbol k)
 \text{ and }  
\dot{\boldsymbol k} = -\, e  \left( {\boldsymbol E}  
+ \dot{\boldsymbol r} \, \cross\, {\boldsymbol B} 
	\right ) \nn
\Rightarrow & \, \dot{\boldsymbol r}  = \mathcal{D}_s ({\boldsymbol k})
	\left[   \boldsymbol{w}_s ({\boldsymbol k}) +
	 e \, {\boldsymbol E}  \cross  
\boldsymbol \Omega^s ({\boldsymbol k})   + 
	 e   \left \lbrace \boldsymbol \Omega^s (\boldsymbol k )\cdot 
 \boldsymbol{w}_s ({\boldsymbol k}) 
 \right \rbrace  \boldsymbol B  \right] 
 \text{ and } 
\dot{\boldsymbol k}  = -\, e \,\mathcal{D}_s ({\boldsymbol k})
  \left[   {\boldsymbol E} 
+   \boldsymbol{w}_s ({\boldsymbol k}) \cross  {\boldsymbol B} 
+  e  \left (  {\boldsymbol E}\cdot  {\boldsymbol B} \right )  
\boldsymbol \Omega^s  ({\boldsymbol k})\right],
\end{align}
where
\begin{align}
\boldsymbol{w}_s ({\boldsymbol k}) = {\boldsymbol  v}^s(\boldsymbol k )
+  {\boldsymbol   v}_{(m)} ({\boldsymbol k} ) \,.
\end{align}
The presence of $\Omega^s $ and $ \mathcal{D}_s $ reflect the nontrivial role played by a nonzero BC, as compared to the scenarios when the BC vanishes (see, for example, the systems discussed in Refs.~\cite{ips-kush, ips_tilted_dirac}). In particular, an extra term in the form of $- \dot{\boldsymbol k} \, \cross \, \boldsymbol \Omega^s $ represents an anomalous velocity, with the BC playing the counterpart of the magnetic field, albeit in the momentum space.

The bare-bones kinetic equation, arising out of the fundamental Boltzmann's transport formalism, is expressed as
\begin{align}
\label{eqkin32}
& \left [
 \partial_t  
+ {\boldsymbol w}_s ({\boldsymbol k}) \cdot \nabla_{\boldsymbol r} 
-e \left \lbrace  \boldsymbol{E}
+  {\boldsymbol w}_s ({\boldsymbol k})  \times {\boldsymbol B} 
\right \rbrace  \cdot 
\nabla_{\boldsymbol k} \right ] f_s (\boldsymbol r, \boldsymbol k, t)
=  I_{\text{coll}} [f_s ({\boldsymbol k}) ]\,,
\end{align}
where $f_s  (\boldsymbol r, \boldsymbol k, t) $ represents the nonequilibrium quasipaticle-distribution function associated with the band $s$. A slight deviation from $ f_0 (\xi_s(\boldsymbol k) ) $ is parametrised by $ \delta f_s  (\boldsymbol r, \boldsymbol k, t) \equiv 
f_s  (\boldsymbol r, \boldsymbol k, t) - f_0 (\xi_s (\boldsymbol k) ) $, caused by the the probe electric field, and is assumed to be of the same order of smallness as $|\boldsymbol E|$ (say, $\delta \varsigma$). Here, we restrict ourselves to spatially-uniform and time-independent electromagnetic fields. Consequently, $f_s $ must must also be independent of position and time, so that $ \delta f_s  (\boldsymbol r, \boldsymbol k, t) = \delta f_s  (\boldsymbol k)$. When we retain terms upto linear-order-in-$ \delta \varsigma $, we restrict ourselves to the so-called linear-response regime. The \textit{collision integral}, $ I_{\text{coll}} [f_s ({\boldsymbol k}) ] $, takes care of the relevant scattering processes instrumental in relaxing $f_s ({\boldsymbol k})$ towards the equilibrium value of $f_0 (\xi_s({\boldsymbol k}))$.

For elastic point-scattering mechanisms, the collision integral takes the form of
\begin{align}
I_{\text{coll}} [f_s({\boldsymbol k})]
 = \sum \limits_{\tilde s}
\int_{k'}
\mathcal{M}_{ s, \tilde s} (\boldsymbol k,\boldsymbol k^\prime)
\left[ f_{\tilde s} ( \boldsymbol k^\prime)  - f_s (\boldsymbol k) \right ] ,
\text{ where  }
\int_k \equiv \int  \frac{ d^3 \boldsymbol k } 
{ (2\,\pi)^3  \, \mathcal{D}_{\tilde s} ({\boldsymbol k} ) }
\end{align}
symbolises the 3d momentum-space integral, incorporating the modified phase-space factor due to the BC.
The scattering rate, not involving any energy-dissipation, is a direct consequence of the Fermi's golden rule, reflected by
\begin{align}
\mathcal{M}_{ s, \tilde s} (\boldsymbol k,\boldsymbol k^\prime) 
= \frac{2\, \pi \, \rho_{\rm imp} } {V} \,
\Big \vert \left \lbrace  \psi_{\tilde s }({ \boldsymbol k^\prime }) 
\right \rbrace^\dagger 
\; {\mathcal V} _{ s, \tilde s} (\boldsymbol k,\boldsymbol k^\prime) 
  \;  \psi_s ({ \boldsymbol k}) \Big \vert^2 \,
 \delta \Big( \xi_{\tilde s } (\boldsymbol k^\prime)- \xi_s (\boldsymbol k ) \Big) \,.
\end{align}
Here, $ \rho_{\rm imp}$ embodies the impurity-concentration (acting as the scattering centres), $V$ denotes the system's spatial volume, and ${\mathcal V} _{ s, \tilde s} (\boldsymbol k,\boldsymbol k^\prime) $ represents the scattering-potential matrix. Restricting to elastic and spinless scatters, $ {\mathcal V} _{ s, \tilde s} (\boldsymbol k,\boldsymbol k^\prime) 
=  \mathbb{I}_{2 \times 2} \, {\mathcal V} _{ s, \tilde s} $, reducing to an identity matrix in the spinor space sans any momentum dependence. Thus, finally we can use the equation,
\begin{align}
\label{eqoverlap0}
\mathcal{M}_{ s, \tilde s} (\boldsymbol k,\boldsymbol k^\prime) 
= \frac{2\, \pi \, \rho_{\rm imp} 
\, |{\mathcal V} _{ s, \tilde s} |^2} 
{V} \,
\Big \vert \left \lbrace  \psi_{\tilde s }({ \boldsymbol k^\prime }) \right \rbrace^\dagger 
\; \psi_{s}({ \boldsymbol k}) \Big \vert^2 \,
 \delta \Big( \xi_{\tilde \chi , \tilde s } (\boldsymbol k^\prime)
 - \xi_{  \chi , s } (\boldsymbol k ) \Big) \,,
\end{align}
assisted by Eq.~\eqref{eqev}. Assuming a symmetric relation between the $ s = \pm 1$ bands, we parametrise the scattering strengths via two parameters, expressed as
\begin{align}
\big |{\mathcal V}_{1, 1} \big |^2  
 =  \big | {\mathcal V}_{-1,- 1} \big|^2
  \equiv  \frac{ 2\,\pi} {\rho_{\rm imp}} \, \beta_{\rm intra}
 \text{ and } 
 \big |{\mathcal V}_{ 1,-1} \big |^2 =   \big |{\mathcal V}_{-1, 1} \big|^2
 \equiv \frac{ 2\,\pi} {\rho_{\rm imp}}\, \beta_{\rm inter} \,.
\end{align} 
Here, the notation is chosen to reflect the fact that $\beta_{\rm intra}$ and $\beta_{\rm inter}$ stand for the strengths of intraband and interband scatterings, respectively.

\subsection{Linearised Boltzmann equation and its solutions}

Putting together all the ingredients of the preceding subsections, we set out to solve the \textit{linearised Boltzmann equation}, captured by
\begin{align}
\label{eqkin5}
& - e\, {\mathcal D}_s ({\boldsymbol k})
 \left [
\left \lbrace {\boldsymbol{w}}_s ({\boldsymbol k})
+ e \,\Big(
{\boldsymbol \Omega}^s ({\boldsymbol k}) 
\cdot {\boldsymbol{w}}_s   ({\boldsymbol k})
 \Big )  \boldsymbol B \right \rbrace
\cdot {\boldsymbol E}  
\; \;	\frac{\partial  f_0 (\xi_s({\boldsymbol k})) }
 {\partial \xi_s({\boldsymbol k}) }
+  
\left \lbrace   {\boldsymbol{w}}_s ({\boldsymbol k})
\cross  {\boldsymbol B}  \right \rbrace
\cdot \nabla_{\boldsymbol k}
\, \delta f_s (\boldsymbol k) \right] 
 =  I_{\text{coll}} [f_s ({\boldsymbol k})] \,.
\end{align}
For the sake of definiteness, we choose the set-up comprising $\boldsymbol B  =  B\, \boldsymbol{\hat{z}}$ and $ \boldsymbol E  =  E\, \boldsymbol{\hat{z}}$.
Parametrising the deviation in particle-distribution as
\begin{align}
\label{eqansatz}
	\delta f_s (\boldsymbol {k}) =
-\,	 e\, 	\frac{\partial  f_0 (\xi_s) } {\partial \xi_s } 
	\,   {\boldsymbol E}  \cdot \bm{\Lambda}_s (\boldsymbol {k} )
=
-\,	 e\, 	\frac{\partial  f_0 (\xi_s({\boldsymbol k})) }
 {\partial \xi_s ({\boldsymbol k})} 
	\,   E \, {\Lambda}^z_s ( \boldsymbol {k} ) \,,
\end{align}
where $ \bm{\Lambda}_s (\boldsymbol {k} ) $ is the vectorial mean-free path, we immediately infer that only the $z$-component of $ \bm{\Lambda}_s ( \boldsymbol {k} ) $ [i.e., $ {\Lambda}^z_s (\boldsymbol {k} )  $] emerges as the nontrivial component to be computed for the purpose of linear response. The relevant equation to be solved is
\begin{align}
\label{eqvec}
&  w^z_s ({\boldsymbol k})
+ e \, B \left [
{\boldsymbol \Omega}^{s} ({\boldsymbol k})
 \cdot {\boldsymbol{w}}_s ({\boldsymbol k})  \right ]
-\, 
e \, B \left [ {\boldsymbol{w}}_s ({\boldsymbol k}) \cross   
\boldsymbol{\hat z}  ({\boldsymbol k}) \right ] 
\cdot \nabla_{\boldsymbol k} {\Lambda}^z_s (\boldsymbol {k} )  
 = {\mathcal D}^{-1}_s  ({\boldsymbol k})
\sum \limits_{\tilde \chi, \tilde s}
\int_{k'}
\mathcal{M}_{ s, \tilde s} (\boldsymbol k,\boldsymbol k^\prime)
\left[ 
 {\Lambda}^z_{\tilde s} (\boldsymbol {k}^\prime ) 
 -  {\Lambda}^z_s (\boldsymbol {k} )  \right ],
\end{align}
for which we put forward the self-consistent ansatz that 
$ {\Lambda}^z_s \equiv {\Lambda}^z_s ( \mu, \theta)$ at an energy $\mu$. In other words, $\delta f_s (\boldsymbol {k})$ can depend only on the polar angle, $\theta$, and the chemical potential, $\mu$, because the integral over the full momentum space effectively gets substituted by an integral over the Fermi surface at energy $\xi_s (\boldsymbol k) = \mu $. Basically, the dependence on the azimuthal angle ($\phi$) drops out because of the remaining rotational symmetry of the system about the $z$-axis. As a result, the momentum-space integrals reduce to the respective Fermi surfaces, $\xi_s 
(k_F^s , \theta )= \mu $, with the $\theta$-dependent radii of $ \lbrace k_F^s (\theta) \rbrace $ (aka the local Fermi momenta). The self-consistency is evident from the fact that $\left [ {\boldsymbol \Omega}^s (\boldsymbol k) \cdot
 {\boldsymbol{w}}_s (\boldsymbol k)  \right ] $ is $\phi$-independent and $ \left [  {\boldsymbol{w}}_s 
 (\boldsymbol k)  \cross   \boldsymbol{\hat z}  \right ] 
\cdot \nabla_{\boldsymbol k} {\Lambda}^z_s (\mu, \theta ) $ reduces to zero.

The inevitable diktat of the above arguments is that, although the spinor overlaps [viz., $\big \vert \left \lbrace  \psi_{\tilde s }({ \boldsymbol k^\prime }) \right \rbrace^\dagger 
\; \psi_{s}({ \boldsymbol k}) \big \vert^2 $ evaluated using Eq.~\eqref{eqev}] contain the azimuthal angles,
$\phi$ and $\phi^\prime $, they will drop out on performing the azimuthal-angle integrations. This allows us the simplicity of using the overlap-function defined as 
\begin{align}
\label{eqoverlap}
{\mathcal T }_{ s, \tilde s} (\theta, \theta^\prime)
& = \frac{ 1 + s \, \tilde s \cos \theta \, \cos \theta^\prime }{2} \, ,
\end{align}
after already performing the trivial $\phi$-integrals. Plugging in the $\phi$-independent forms, Eq.~\eqref{eqvec} reduces to
\begin{align}
\label{eq_lambda_mu}
h_s (\mu, \theta) = 
 \sum_{\tilde s} V  
\int_{k^\prime }\, \mathcal{M}_{ s, \tilde s} 
(\boldsymbol k,\boldsymbol k^\prime)
\, {\Lambda}^z_{\tilde s} ( \mu, \theta^\prime ) 
 - \frac{ {\Lambda}^z_s  (\mu, \theta)} 
{\tau_s(\mu, \theta)}  \,,
\end{align}
where
\begin{align}
\label{def-tau-h}
\tau^{-1}_s(\mu, \theta)
= \sum_{\tilde s} V \int_{k^\prime }
\mathcal{M}_{ s, \tilde s} (\boldsymbol k,\boldsymbol k^\prime) \,, \quad
h_s (\mu, \theta) = {\mathcal D}_s ({\boldsymbol k}) \left[ w^z_s ({\boldsymbol k})
+ e \, B \left \lbrace
{\boldsymbol \Omega}^s ({\boldsymbol k}) \cdot {\boldsymbol{w}}_s ({\boldsymbol k})
  \right \rbrace \right].	
\end{align}
Working with the integrals reduced to the respective Fermi surfaces at energy $ \mu $, we need to solve
\begin{align}
\label{eqlambdamu}
& h_s(\mu, \theta) + \frac{ {\Lambda}^z_s  (\mu, \theta)} 
{\tau_s(\mu, \theta)} =
\sum_{\tilde s}  
\frac{ \rho_{\rm imp} 
\, |{\mathcal V} _{ s, \tilde s} |^2 }
{ 4\, \pi }
\int d\theta^\prime \, \frac{\sin \theta^\prime \left (k^\prime \right )^3 
\, {\mathcal D}^{-1}_{\tilde s} ({\boldsymbol k}^\prime)}
{  |\boldsymbol k^\prime \cdot {\boldsymbol{w}}_s (\boldsymbol k^\prime)  | }
\, {\mathcal T }_{ s, \tilde s} (\theta, \theta^\prime)
\, {\Lambda}^z_{\tilde s} (\mu, \theta^\prime ) 
\Big \vert_{ k^\prime = k_F^{\tilde s} } \,.
\end{align}
In the above expression, the factor $ \left (k^\prime \right )^2  \sin \theta^\prime $ arises as the Jacobian for employing the spherical-polar coordinates, and the part $$  \big |\boldsymbol {\hat k^\prime} \cdot 
 \nabla_{\boldsymbol k^\prime} \xi_s({\boldsymbol k^\prime}) \big |^{-1}  
 = k^\prime / |\boldsymbol k^\prime \cdot {\boldsymbol{w}}_s (\boldsymbol k^\prime)  |$$
originates from converting $ \delta \Big( \xi_{\tilde s } (\boldsymbol k^\prime) -\mu \Big) $ to
$\delta (k^\prime - k_F^{\tilde s})$.

As a final step, we observe the form of the wavevectors and their overlaps [cf. Eqs.~\eqref{eqoverlap0} and \eqref{eqoverlap}],
to formulate the solutions as
\begin{align}
\label{eqLamsol}
 {\Lambda}^z_s (\mu,\theta ) & = 
\tau_s (\mu,\theta) \left [ \lambda_s - h_{s} (\mu, \theta) 
 + a_s \, \cos \theta  \right ],
\end{align}
comprising 4 undetermined coefficients, $\lbrace \lambda_s , \, a_s  \rbrace $. Using the four linear equations furnished by Eq.~\eqref{eqlambdamu}, this translates into the problem of solving the following matrix equation: 
\begin{align}
\label{eqmatrix}
\mathcal A \, \mathcal N = \mathcal G \,, \text{ where }
\mathcal N =\begin{bmatrix}
\lambda_1 & a_1  & \lambda_{-1} & a_{-1}
\end{bmatrix}^{\rm T}\,.
\end{align}
Interestingly, the above equation is not sufficient to obtain the complete set of solutions, because $ \mathcal A $'s rank is lower than its dimension. Thankfully, there is the electron-number conservation to take care of this issue, furnishing the extra equation of
\begin{align}
\label{eqcon}
\sum \limits_{s}	\int_k  \delta f_s (\boldsymbol {k}) = 0\,.
\end{align}

\begin{figure*}[]
\subfigure[]{\includegraphics[width=  \textwidth]{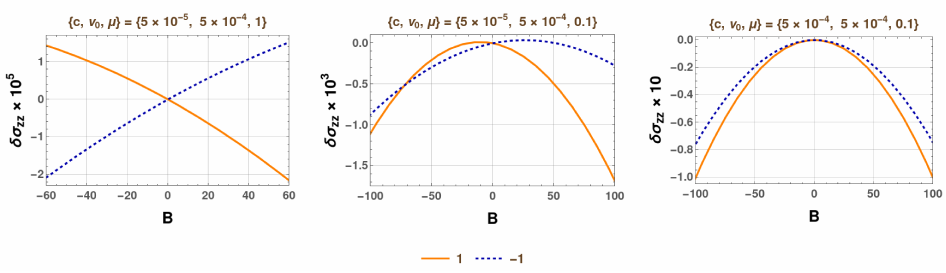}}
\subfigure[]{\includegraphics[width=  \textwidth]{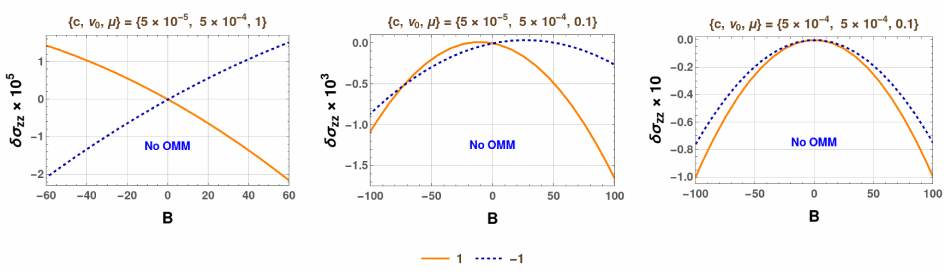}}
\caption{\label{figsep}$\delta \sigma_{zz} (s) $ from each of the bands considering no interband interactions, with $\beta_{\rm intra}  $ set to unity: While subfigure (a) depicts the variation of the full conductivity with $B$ (in eV$^2$) when OMM is taken into account appropriately, subfigure (b) represents the conductivity versus $B$ characteristics when OMM is neglected. The number in each plot-legend indicates the index of the band (i.e., $1$ or $-1$). The three frames in each subfigure represent three distinct sets of parameter values, as indicated in the plot-labels, with $c$ in eV$^{-1}$, $v_0 $ being unitless, and $\mu $ in eV.}
\end{figure*}

Once we obtain the solutions, the charge-current density along the $z$-direction turns out to be
\begin{align}
\label{def_cur}
	J_z^{\rm tot} =- \,\frac{e}{ V }\sum_s
\int_k	(\dot{\boldsymbol {r}} \cdot \boldsymbol{\hat z})
\;  \delta f_s (\boldsymbol k)\,,
\end{align}
resulting in the longitudinal magnetoconductivity of
\begin{align}
\label{def_sig_long}
\sigma_{zz}^{\rm tot} = \sum_s \sigma_{zz}^s \,, \quad
 \sigma_{zz}^s
=  -\,\frac{e^2 } { V } 
\int \frac{d^3 {\boldsymbol k}} {(2\, \pi)^3} 
	\left[   w^z_s (\boldsymbol k)
+ e \, B  \left \lbrace \boldsymbol \Omega^s (\boldsymbol k)
\cdot 	  \boldsymbol{w}_s (\boldsymbol k) \right  \rbrace   \right] 
\delta \big (\xi_s (\boldsymbol k)-\mu \big) \, {\Lambda}^z_s (\mu, \theta )	\,. 
\end{align}

\subsection{Results and discussions} 
\label{secres}

Through numerical computations, we solve Eq.~\eqref{eqlambdamu} to generate representative values which shed light on the nature of the emergent response. In particular, we illustrate the behaviour of
$$\delta \sigma_{zz} (s) \equiv \sigma^s_{zz} (B) / \sigma^s_{zz}(B=0)  -1 ,$$
where we subtract off the residual conductivity at $B=0$ (representative of the Drude part). Our choice of parameter regimes is dictated by the values mentioned in the literature \cite{long-ksm}.

We start with the simplest case when there is no interband scattering, i.e., $\beta_{\rm inter } $ is set to zero. This leads to the reduction of $  \mathcal A $ into the direct sum of the two matrices,  $ \mathcal{A}_1 $ and $ \mathcal{A}_2 $, where each is a $2$-dimensional square matrix. In every case, we find that they have rank $1$ (instead of $2$) and, hence, needs to be supplemented by the conservation equation, $ \int_k  \delta f_s (\boldsymbol {k}) = 0$, to determine all the unknown coefficients. Fig.~\ref{figsep} captures such a scenario. We find that the results are not changed much on disregarding the OMM-contributions (see the bottom panel). Overall, the asymmetry of the curves about the vertical axis betrays the fact that there is a coexistence of linear-in-$B$ and quadratic-in-$B$ terms in all the response-curves. In each case, we observe a parabolic nature, with the parabola \textit{always} curving downwards. One can check that this is a generic feature when a single band at a single node participates in transport. In Ref.~\cite{sharma-2023}, this was discussed for WSMs, when there is no internode scattering and, hence, only \textit{one} band at a node has to be considered. There, as soon as internode scattering ($\beta_{2node}$) was introduced, the curves are seen to flip when $\beta_{2node}$ is below a critical value. The fundamental reason is rightly attributed to the charge conservation equation that needs to be used: while for a single node, charge conservation must be satisfied individually at each node, a nonzero internode-coupling demands that the global charge conservation from summing over the two nodes must be the relevant equation.

\begin{figure*}[]
\subfigure[]{\includegraphics[width=  \textwidth]{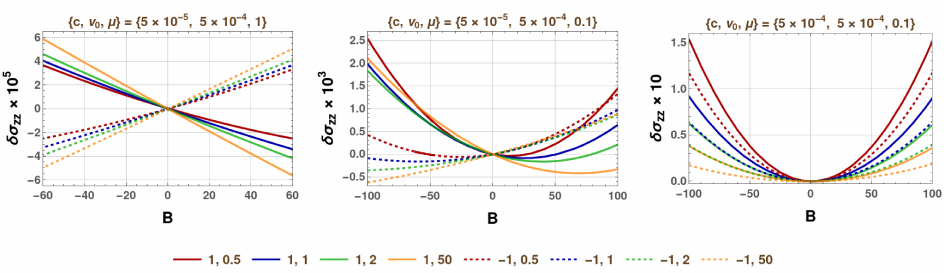}}
\subfigure[]{\includegraphics[width=  \textwidth]{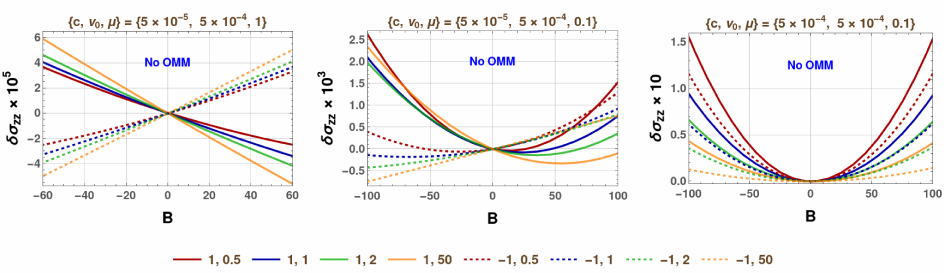}}
\caption{\label{figall}$\delta \sigma_{zz} (s) $ from each of the bands considering both intraband and interband interactions, with $\beta_{\rm intra}  $ set to unity: While subfigure (a) depicts the variation of the full conductivity with $B$ (in eV$^2$) when OMM is taken into account appropriately, subfigure (b) represents the conductivity versus $B$ profiles when OMM is neglected. The two numbers in each plot-legend indicate the index of the band (i.e., $1$ or $-1$) and the value of $\beta_{\rm inter}  / \beta_{\rm intra} $, respectively. The three frames in each subfigure represent three distinct sets of parameter values, as indicated in the plot-labels, with $c$ in eV$^{-1}$, $v_0 $ being unitless, and $\mu $ in eV.}
\end{figure*}

Fig.~\ref{figall} shows the generic situations when the quasiparticles populating both the bands can scatter with each other. The $4$-dimensional square matrix $ \mathcal{A}$ is found to have rank $3$, ensuring the internal consistency of the available constraints. This is because the charge-conservation constraint imposes another equation [viz., Eq.~\eqref{eqcon}], providing the complete set of linearly-independent equations essential to determine the unknown coefficients \cite{timm, ips-exact-spin1}. Setting $\beta_{\rm intra}  = 1$, we vary the ratio $\beta_{\rm inter}  / \beta_{\rm intra} $ over a wide range of values. The qualitative characteristics of the curves are not drastically altered by such variations, other than making them flatter and flatter. One remarkably striking difference that emerges, as soon as we set $\beta_{\rm inter}$ to nonzero, is that the parabolas now curve upwards. Therefore, for multi-Fermi-surface nodal points, a nonzero interband coupling brings about the same effects as a nonzero internode coupling does for chirally-conjugate nodes \cite{sharma-2023, ips-exact-spin1}. This is again tied up with the number (and form) of charge-conservation equations that one must use, depending on the number of bands which are intertwined with each other in terms of allowing particle-flow amongst themselves.

The presence of a combination of linear-in-$B$ and quadratic-in-$B$ terms are apparent from the shape of the curves. In some regimes, the linear-$B$ behaviour dominates [cf. leftmost panel] and, in some other, the $B^2$-part takes over [cf. rightmost panel]. In order to get an idea regarding the significance of the OMM, we also plot the conductivity by artificially switching off the OMM-parts. We observe that the OMM changes the magnitude of the response-curves very minutely. Hence, unlike the OMM-effects in other semimetallic systems with band-inversion Weyl-like behaviour \cite{ips-ruiz, ips-rsw-ph, ips-spin1-ph, ips-exact-spin1}, here their influence is minuscule.

When a system is subjected to homogeneous external fields, in the absence of any other scale in the problem, the Onsager-Casimir reciprocity relation must hold \cite{onsager31_reciprocal, onsager2, onsager3}, which says that $\sigma_{zz} (\boldsymbol {B})= \sigma_{zz} (- \boldsymbol {B})$. This forbids any linear-in-$ B$ term in the response, unless the change of sign of $ \boldsymbol {B} $ is compensated by a simultaneous change of sign in another parameter of the system. A nonzero SMM provides us exactly with such a compensating sign, such that the relation $\sigma_{zz} (B_z, s)= \sigma_{zz} ( - B_z, - s)$ is satisfied in our set-up, thus fulfilling the Onsager-Casimir constraints~\cite{cortijo16_linear}. This allows the linear-in-$B$ terms to appear, which we clearly see from our explicit results. We contrast this with the case when we have a pair of chirally-conjugate nodes possessing rotational symmetry (i.e., for example, without having any anisotropy such as tilt). One might argue that the possibility of having $\sigma_{zz} (B_z, \chi )= \sigma_{zz} ( - B_z, - \chi )$ could make a linear-in-$B$ term possible. But the issue is that, the BC-contributions and the band-velocity contributions from the OMM appear as factors proportional to $\chi \, \boldsymbol k \cdot \boldsymbol B /k^n = \chi \, k_z \, B_z /k^n $, with a surviving $k_z $ factor from $\boldsymbol v^s $ (i.e., the zeroth order $\boldsymbol B $-independent band-velocity) giving an odd-in-$k_z$ integral evaluating to zero. In other words, the answers there exclusively contain terms like $\chi^ 2\, B^2 = B^2 $, thus preventing a linear term. This is what was observed in Ref.~\cite{ips-exact-spin1}, where we considered pseudospin-1 nodes with quadratic-in-momentum corrections.

For the benefit of the reader, let us elaborate on the precise origins of the linear-in-$B$ and quadratic-in-$B$ terms. From Eqs.~\eqref{def_sig_long}, \eqref{def-tau-h}, and \eqref{eqLamsol}, we find that
\begin{align}
\label{def_sig_long2}
 &- V\,\sigma_{zz}^s/e^2 \nn
& = \int \frac{d^3 {\boldsymbol k}} {(2\, \pi)^3} 
\,\tau_s (\mu,\theta) 
	\left[   w^z_s (\boldsymbol k)
+ e \, B  \left \lbrace \boldsymbol \Omega^s (\boldsymbol k)
\cdot 	  \boldsymbol{w}_s (\boldsymbol k) \right  \rbrace   \right] 
\left [ \lambda_s - 
{\mathcal D}_s ({\boldsymbol k}) \left \lbrace
 w^z_s ({\boldsymbol k})
+ e \, B \Big (
{\boldsymbol \Omega}^s ({\boldsymbol k}) \cdot {\boldsymbol{w}}_s ({\boldsymbol k})
  \Big ) \right \rbrace
 + \frac{a_s \, k_z} {k}  \right ] \delta \big (\xi_s (\boldsymbol k)-\mu \big) 	\,. 
\end{align}
For gaining the requisite physical insight into the scaling of $B$ in the results, it suffices to assume $\tau $ to be ${\boldsymbol k} $-independent, and expand ${\mathcal D}_s ({\boldsymbol k})$ as well as $\delta \big (\xi_s (\boldsymbol k)-\mu \big)$ up to $\mathcal{O} (B^2)$. Now, we have to remember that $\xi_s ({ \boldsymbol k}) =
c\,  k^2 + s \,v_0 \,  k  +  \frac{  e\, v_0 \, k_z\, B} {2 \, k^2 } 
+  \frac{ e \,g\,\mu_B \,s \, B} {2}  $ and
$ {\boldsymbol  w}_s ( \boldsymbol k) 
=  \frac{ 2 \, c\, k + s\, v_0  } {  k }  
\left \lbrace k_x,\,k_y, \, k_z \right \rbrace 
-\frac{e \,v_0\, B } {k_F^4} \left \lbrace
k_z \, k_z , \, k_y\, k_z, \, k_z^2 -\frac{k_F^2}{2}  \right \rbrace  $. The only term in the integrand which can give rise to a nonzero linear-in-$B$ term originates from the part of the integrand proportional to
$\left[  {\boldsymbol  v}^s(\boldsymbol k )\right]^2 {\boldsymbol B} \cdot {\boldsymbol S}_s \,\delta^\prime  \big ( {\tilde \varepsilon}_s  ({ \boldsymbol k}) -\mu \big )$, because all terms odd in ${\boldsymbol  k}$ vanishes on integration over ${\boldsymbol  k}$.
The remaining even-in-${\boldsymbol  k}$ parts of the integrand comprise $B^2$. This is how the Onsager-Casimir reciprocity relation is satisfied as well, as outlined in the preceding paragraph --- the presence of a nonzero SMM is solely responsible for giving rise to a term $\propto B $.

\section{Comparison with results obtained from relaxation-time approximation}
\label{secsig}

In this section, we employ the simplistic application of the relaxation-time approximation to compute the conductivity for individual bands, which we have used in our earlier works \cite{ips_rahul_ph_strain, rahul-jpcm, ips-ruiz, ips-rsw-ph, ips-shreya, ips-tilted, ips-internode, ips-spin1-ph}. In order to obtain closed-form analytical expressions, we have to expand the $B $-dependent terms upto a given order in $B$, assuming it has a small magnitude, which is anyway required to justify ignoring Landau-level splitting. With this in mind, the following two quantities are expanded as shown below:
\begin{align}
& f_0^\prime \left(\xi_s\right) = 
f_0^\prime \big(  {\tilde \varepsilon}_s ({ \boldsymbol k}) \big)
+ \varepsilon^{(m)} \, f_0^{\prime \prime } \big ( {\tilde \varepsilon}_s ({ \boldsymbol k})\big )
+ \frac{1}{2} \left[ \varepsilon^{(m)} \right ]^2 
f_0^{\prime \prime \prime}\big ( {\tilde \varepsilon}_s  ({ \boldsymbol k})\big )
 + \mathcal{O} (B^3 )\,, 
\text{ where }
 {\tilde \varepsilon}_s  ({ \boldsymbol k})=
 \varepsilon^s  ({ \boldsymbol k}) +  \frac{ e \,g\,\mu_B \,s \, B} {2}\,,
\nn & \text{ and } \mathcal{D}_s = \sum \limits_{n=0}^{2} 
\left [ -e \,  \Omega^z_s \, B \right ]^n 
+ \mathcal{O} (B^3 )\,.
\end{align}
Here, the ``prime'' symbol denotes the operation of partial-differentiation, with respect to the variable shown explicitly within the brackets [e.g., $ f_0^\prime (\varepsilon) \equiv \partial_\varepsilon f_0 (\varepsilon)$]. 
Since we are working in the $T \rightarrow 0 $ limit, we have to use $f_0^\prime (\varepsilon) \rightarrow -\, 
\delta ( \varepsilon - \mu )$. Observing that the radial part of the integrals is with respect to the variable $k$, we need to use the following forms of the concerned expressions:
\begin{align}
& f_0^{\prime \prime } \big ( {\tilde \varepsilon}_s  ({ \boldsymbol k})\big )
= \frac{ \partial_k f_0^\prime \big ( {\tilde \varepsilon}_s  ({ \boldsymbol k})\big ) }
{2 \, c \, k  + s \, v_0 } \,, \quad
f_0^{\prime \prime \prime} \big ( {\tilde \varepsilon}_s  ({ \boldsymbol k})\big )
= \frac{ \partial^2_k f_0^\prime \big ( {\tilde \varepsilon}_s  ({ \boldsymbol k})\big )
}  {v_0^3 \, (2 \, c\, k + s\,  v_0 )^2 }
-
\frac{2 \,c\, \partial_k f_0^\prime \big ( {\tilde \varepsilon}_s  ({ \boldsymbol k})\big )
}  { v_0^3 \, (2 \, c\, k + s\,v_0 )^3} \,.
\end{align}
We refer to Refs.~\cite{ips-kush-review, ips_rahul_ph_strain, ips-rsw-ph} for a review of the generic expressions for the in-plane (i.e., coplanar with the applied electromagnetic fields) components the conductivity tensor.
Noting that the in-plane anomalous-Hall part vanishes \cite{ips-ruiz, ips-rsw-ph, ips-spin1-ph}, the generic expression therein can be expressed as
\begin{align}
\label{eqsig}
   \sigma^s_{i j}
&= - \,e^2 \, \tau  
\int \frac{ d^3 \boldsymbol k}{(2\, \pi)^3 } \, \mathcal{D}_s
\left[  (w_s)_i \, + (W_s)_i \right ]
\left [ (w_s)_j \, + (W_s)_j \right] \,
f^\prime_0 \big (\xi_s \big )  \,,
\text{ where }
\boldsymbol{W}_s 
= e \left  ( {\boldsymbol w}_s \cdot 
  \boldsymbol {\Omega}^s \right  ) \boldsymbol{B}\,.
\end{align}

For our KWN system, only the longitudinal component survives, leading to
\begin{align}
\label{eqrelax}
 & \sigma^s_{zz} = \frac{e^2 \, \tau } {8 \,\pi ^3} 
 \left (
\sigma_{zz}^{0,s} + \sigma_{zz}^{{\rm bc},s}
+ \sigma_{zz}^{{\rm m},s} \right) ,
\nn & \sigma_{zz}^{0,s}  =
\frac{v_0^2 \left(\sqrt{v_0}
-\sqrt{\frac{4 \,c \,{\tilde \mu} }{v_0}+v_0}\right)^2
 \left[\sqrt{\frac{4\, c\, {\tilde \mu} } {v_0}+v_0}
 +(s-1) \,  \sqrt{v_0} \right]^2}
 {6 \,c^2 \,\sqrt{4 \,c\, {\tilde \mu} +v_0^2}}\,,
\nn & \sigma_{zz}^{{\rm bc},s}  =
\frac{2 \, e^2 \,c^2 \, B^2 }  
 {15 \,v_0 \,\sqrt{4 \,c\, {\tilde \mu} + v_0^2}}
\frac{  
2 \,(80\, s-77)\, v_0^{\frac{3}{2}}\,
\sqrt{\frac{4 \,c \,{\tilde \mu} } {v_0} + v_0}
+ 4 \,c \,{\tilde \mu}  \, (63-20 \, s) + (177-160 \,s)\, v_0^2 }
   { \left(\sqrt{v_0}-\sqrt{\frac{4 \, c  \,{\tilde \mu} }   { v_0} + v_0}
 \right)^2}\,,
\nn & \sigma_{zz}^{{\rm m},s}  =
\frac{2  \,e^2 \, c^2 }
{15 \,\sqrt{v_0} \;\sqrt{\frac{4 \,c \,{\tilde \mu} }{v_0} + v_0 } 
\left(4 \,c \,{\tilde \mu} +v_0^2\right)^{5/2}
   \left(\sqrt{v_0}-\sqrt{\frac{4 \,c \,{\tilde \mu} }{v_0}+v_0}\right)^2}
\nn & \hspace{ 1.2 cm } \times   \Bigg [
 128 \,c^3 \,{\tilde \mu}^3 \,(9 \,s-10) + 16 \,c^2\, {\tilde \mu} ^2\, (136\, s-133) \, v_0^2 
+ 216\, c^2 \,{\tilde \mu} ^2\, (5-6\, s)\, v_0^{\frac{3}{2}}
 \,\sqrt{\frac{4 \,c \,{\tilde \mu} }{v_0} + v_0}
  + (118\, s-111) \,v_0^6
 \nn & \hspace{1.75 cm}
 + 2\, (61-70 \,s ) \,v_0^{\frac{11}{2}}
 \, \sqrt{\frac{4 \,c \,{\tilde \mu} }{v_0}+v_0}  
 + 16 \,c \,{\tilde \mu}\,  (59 \,s-56) \,v_0^4
 +2\, c\, {\tilde \mu} \, (361-424 \,s)\,
   v_0^{\frac{7}{2}}\, \sqrt{\frac{4 \,c \,{\tilde \mu} }{v_0}+v_0} \,
 \Bigg ]  \,,
\nn & \text{where } \tilde \mu = \mu - \frac{ e \,g\,\mu_B \,s \, B} {2} \,.
\end{align}
Here, the superscipts ``bc'' and ``m'' stand for the BC-only and OMM-sourced contributions, respectively. We note that $ \sigma_{zz}^{{\rm dr},s} \equiv \sigma_{zz}^{{\rm 0},s} \vert_{B=0}$ represents the Drude part (i.e., the residual conductivity in the absence of a magnetic field).

\begin{figure*}
\includegraphics[width= 0.7 \textwidth]{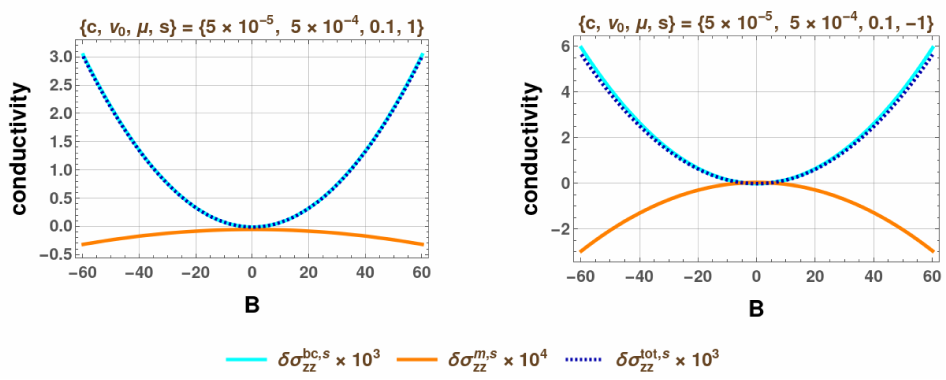}
\caption{\label{figrel}Longiudinal conductivity from the relaxation-time approximation [cf. Eqs.~\eqref{eqrelax} and \eqref{eqrelax2}] for $s=1 $ (left frame) and $s=-1$ (right frame). The parameter values, as indicated in the plot-labels, with $c$ in eV$^{-1}$, $v_0 $ being unitless, and $\mu $ in eV.}
\end{figure*}

For both the bands, the signs of $ \sigma_{zz}^{{\rm bc},s} $ and $\sigma_{zz}^{{\rm m},s}$ are opposite. In the parameter regimes considered here, $ - \,\sigma_{zz}^{{\rm m},s} < \sigma_{zz}^{{\rm bc},s}$, with the overall response always remaining positive for both the bands. A glimpse of the resulting behaviour is captured in Fig.~\ref{figrel}, dissecting the contributions from the two sources. The functions plotted are defined as:
\begin{align}
\label{eqrelax2}
\delta \sigma_{zz}^{{\rm bc},s} \equiv 
\frac{ \sigma_{zz}^{{\rm bc},s} +  \sigma_{zz}^{0,s} } 
{\sigma_{zz}^{{\rm dr},s}} -1 \,, \quad
\delta \sigma_{zz}^{{\rm m},s} \equiv 
\frac{ \sigma_{zz}^{{\rm m},s} + \sigma_{zz}^{0,s}  }
 {\sigma_{zz}^{{\rm dr},s}}   -1\, \,, \quad
\delta \sigma_{zz}^{{\rm tot},s} \equiv 
\frac{ \sigma_{zz}^{{\rm bc},s} + \sigma_{zz}^{{\rm m},s} + \sigma_{zz}^{0,s}  }
 {\sigma_{zz}^{{\rm dr},s}}   -1\,.
\end{align}
The limitations of the relaxation-time approximation are apparent from comparing with the exact results obtained in the earlier section. The use of a momentum-independent $ \tau $ strips off any angular dependence from it, which endows the curves with positive values, monotonically increasing with $| B| $. Incorporating a proper $\theta$-dependent $\tau$, depending on the nature of the impurity-scattering potential we might be interested in (see, for example, Refs.~\cite{titus-ksm, ips-kush}), can potentially remedy this deficiency. But we refrain from doing so here, since we have already obtained the exact answers by fully solving the Boltzmann equations.

\section{Summary, discussions, and future perspectives}
\label{secsum}

In summary, we have systematically computed the longitudinal magnetoconductivety contributed by an isolated KWN of isotropic nature. The most intriguing aspect of this system is that, at $\mu>0$, the analogue of the ``valence band'' of an archetypal WSM gets promoted to a positive-energy band. Consequently, it no longer remains a spectator band, unlike for the WSMs. The picture derived from our detailed analysis is pretty complete, because tilting is prohibited in the spectrum. We have taken care to include all sources that can couple to $\boldsymbol B $ and, thus, affect the overall response. This includes the BC, the OMM, and the SMM, all of which have been treated at an equal footing. A remarkable feature of the nonzero SMM is that it gives rise to nonzero linear-in-$ B $ terms in the conductivity, satisfying the Onsager-Casimir reciprocity relations. As such, the importance of including the effects of the SMM cannot be overemphasised. However, the OMM plays only a minor role. Solving the Boltzmann equations in their full glory has enabled us to pinpoint the incredibly strong dependence on the values of the intraband- and interband-scattering strengths. Dissecting the features for zero and nonzero interband scatterings, we have unravelled the intriguing fact that a finite $\beta_{\rm inter}$ fundamentally changes the fermionic distributions functions. The upshot is that the conductivity curves can flip in sign compared to the $\beta_{\rm inter} = 0 $ situation.

An immediate direction to pursue is to repeat our calculations for a magnetic field which is not totally collinear with the electric field \cite{girish2023}. Another straightforward but laborious task comprises computing the transmission charactertistics in tunneling problems, figuring out what new features the quadratic-in-momentum-dominated spectra of the KWNs can bring about. This is in the same spirit as studied in some of our earlier works for other nonlinear-in-momentum spectra \cite{fang, zhu, ips3by2, deng2020, ips-aritra, ips-jns}. Such a nonlinearity in the propagation-direction of the quasiparticles inevitably necessitates the inclusion of evanescent waves. Coupled with the double-band contributions, this will proliferate the unknown coefficients to be solved for \cite{banerjee, ips-abs-semid, deng2020, ips-aritra, ips-jns, ips_tunnel_qbcp_corr, ips_tunnel_qbcp_delta}. Nevertheless, it will be a rewarding exercise to take up that challenge.
Last but not the least, one needs to look at finite temperatures \cite{ips_rahul_ph_strain, rahul-jpcm, ips-ruiz, ips-shreya, ips-rsw-ph, ips_tilted_dirac}, when one can also determine the other linear-response coefficients (namely the thermoelectric and thermal-conductivity tensors) in unison, using the same structure of Boltzmann equations.

 
\bibliography{ref_ksm}


\end{document}